\documentstyle[a4,12pt]{article}
\begin{document}
\title{
Kinetics of macrosteps under diffusion and thermal interactions in
stagnant media}

\author{Serge Yu. Potapenko}
\date{01/20/94}
\maketitle

\centerline{Institute of Applied Physics, Russian
Academy of Sciences}
\centerline{46 Ul`anov street, 603600 Nizhny Novgorod,
Russian Federation}

\begin{abstract}
The step motions considered are those in which crystallization is
controlled by a single diffusion process, either the substance diffusion
for growth from solution or the flow of latent heat from the step for
growth from melt. Quasi-static diffusion nearby two parallel steps and
nearby a train of parallel steps is considered.  Terraces and step
risers are assumed flat with arbitrary angles between them. Flux of the
crystallizing substance to the crystal or the heat sources on the
crystal surface are nonvanishing on the step risers only. Kinematics of
the steps under the diffusion interaction is investigated, when
velocities of the steps are controlled by the supersaturation at the
feet of the steps.

For the solution growth it is shown that an equidistant train of steps
is unstable against to doubling of period, i.e. neighbouring steps are
attracted, when the angle of the riser is more than $\pi/3$.  The
diffusion interaction stabilizes the train of gently sloping steps.
For the melt growth at equal thermal conductivities of the crystal and
of the melt, the thermal interaction stabilizes the train of steps
without overhanging when the riser angle is less than $\pi/2$. If the
heat transfers through the crystals, then the equidistant train is
stable at all riser angles.

\end{abstract}

\section{Inroduction}
When the layered crystal growth is treated, the problem of stability of
the moving steps with respect to lateral perturabations and
longitudinal displacement of the steps from steady state positions
arises.  Bunching of the steps in the second kind of instability often
leads to macrostep formation. Kinematic waves of the step density were
investigated in refs.~\cite{F,Ch}. A model describing the step train
evolution \cite{BG,BR} was studied under proposing that the velocities
of the steps is a function depending on the spacing between them. In
particular it was found that the perturbations of the spacings of the
equidistant train are of merely oscillating character for the identical
dependences on the widths of the adjacent upper and lower terraces.

The foundation of the diffusion instability studies was started in the
famous papers by Mullins and Sekerka \cite{ms}. The instabilities of
the steps on layer-grown crystal surface, when the step motion is
controlled by the surface diffusion of crystallizing substance, were
investigated in refs. \cite{Sch,Bales}. In these works a possible
asymmetry in attachment of adatoms to the step from the upper and lower
terraces was taken into account. Stability analysis which takes account
of the inluence of the step motion on the diffusion fields, i.e. without
quasi-static approximation, has recently been carried out \cite{Liu}.
It was shown that the equidistant step train is stable under
longitudinal fluctuation in the distances between steps in contrast to
the model \cite{BG,BR}.

During crystal growth from solution and from melt, the diffusion of
substance in the first case and latent heat in the second case is,
generally speaking, a of volume character. A model of stricly
equidistant train of the step was considered in ref. \cite{Ch}, where
the steps were imagined as semicylinders. A more realistic model of
pair of parallel straigth steps with flat risers and treads was
considered in ref.  \cite{Nish}. However, the conformal mapping used
there allows the diffusion field to be calculated only for definite
relations between the step parameters and their separation.

In sections 3 and 4 of the present paper, the volume diffusion
during macrosteps motion for growth from solution is considered within
the quasi-static approximation. In section 5, the longitudinal
stability of the equidistant macrostep train is treated under the
assumption that, the step motion is governed by the supersaturation at
the step foot.  The heat transfer for growth from melt and the
stability of the train are studied in section 6.

\section{Formulation of the diffusion problems}
Let us consider quasi-static diffusion, when the influence of the step
motion on the diffusion field can be neglected. To do this the step
velocity $V$ must be small enough, i.e., $V L\ll D$, where $L$ is the
step separation and $D$ stands for the diffusivity. Assuming that the
distances between the steps are small compared to the thickness of the
diffusion layer in liquid phase we can to neglect the motion of the
liquid and describe the system by the diffusion equation for a
stagnant medium. We shall consider the steps as being straight and
parallel, which reduces the problem to solving a two-dimensional
diffusion equation for the solute concentration $C$;
\begin{equation}
{\partial^2 C\over\partial x^2} + {\partial^2 C\over\partial y^2} =0.
\end{equation}
We shall assume that the normal component of the solute flux to the
crystal surface is equal to zero at terraces and to some constants at
the step risers. Thus, mathematically, one has to find a harmonic
function, which is analytic in space occupied by the solution and has
a definite normal derivative at the boundary.

This way of problem statement permits one to describe the diffusion
processes for a step with a large number of the elementary layers.
The problem for elementary steps will be discussed in the
Conclusion.

Further, to solve the diffusion problems we shall use the conformal
transformations of the coordinates $z\rightarrow w,\; z=x+i y\in Z,\;
w=u+i v\in W$, mapping the area of $Z$-plain occupied by the solution
into the upper half area of the $W$-plane. The solution of the boundary
problem one can obtain by the Schwarz's integral. In the initial
cordinates, the expression for the solute concentration at the point $z$
through the values of the normal derivative of the concentration
$g(z)$ on the crystal surface $\cal G$ has the form
\begin{equation}
C(z) = {1\over\pi}
\int_{-\infty}^{\infty} d\tau\; \left |{dz\over dw}\right | \;
g[z(\tau)]\; \ln\left |\tau -w(z)\right |\;.
\label{C}
\end{equation}
This boundary problem arises from assuming that the diffusion transfer
of crystallizing substance takes place only in the solution, i.e. the
diffusion coefficient in the crystal is equal to zero. For growth
from melt, the step motion is governed by the latent heat transfer, the
heat may transfering through both the melt and the crystal. As a
rule, the thermal diffusion constants of the crystal $K_c$ and of the
melt $K_m$ are close. In order to simplify the problem, we shall put
these coefficients equal $K_m =K_c\equiv K$. The field of the
undercooling $T_m - T(z)$, where $T_m$ is the melting temperature, in
the unbounded medium with the rate $R$ at which heat is generated on
the crystal surface $\cal G$ is expressed by Green's function of
the two-dimensional problem as
\begin{equation}
T_m - T(z) ={1\over 2\pi K}\int_{\cal G}
|\hbox{d}z'|\;R(z')\,\ln |z -z'|.
\label{T}
\end{equation}

\section{Diffusion nearby two steps}
We shall assume that the steps are straight and parallel to each
other and they have profiles formed by rectilinear segments. Let us
first consider a pair of steps with a separation $L$ and
heights $H_1$ and $H_2$ as is shown in fig.~1. The angles between
the terraces and the risers for these steps are equal to $\pi\alpha_1$
and $\pi\alpha_2$ respectively. Let $\alpha$ lies in the interval
$0<\alpha_{1,2}<1$. The negative values of the angle correspond
to the step moved in the opposite direction. At $1/2<\alpha_{1,2}<1$,
the step has an overhanging peak. Using the Schwarz--Christoffel's
integral
\begin{equation}
z=\int_0^w dw'\;\Gamma (w')\;,
\label{map}
\end{equation}
where $\Gamma =\Gamma_2 (w)$,
\begin{equation}
\Gamma_2 (w) =\left ({w+H_2^*\over w}\right )^{\alpha_2}
\left ({w-L^*\over w-L^* - H_1^*}\right )^{\alpha_1} ,
\label{G2}
\end{equation}
we transfer the area occupied by the solution in the $Z$-plain into the
upper half of the $W$-plain. The parameters $L^*$, $H_1^*$ and $H_2^*$
are determined from the conditions of equality of the distances between
points $A_1$, $A_2$, $B_1$, $B_2$ and the initial values $H_1$,
$L$, $H_2$:
\begin{eqnarray}
L&=&\int_0^{L^*} dw
\left ({w+H_2^*\over w}\right )^{\alpha_2}
\left ({L^* -w\over L^* + H_1^* -w}
\right )^{\alpha_1} ,
\label{L}\\
{H_1\over\sin\pi\alpha_1}  &=&\int_0^{H_1^*} dw
\left ({L^* +H_2^* +w\over L^* +w}\right )^{\alpha_2}
\left ({w\over H_2^* -w}\right )^{\alpha_1} ,
\label{H2}\\
{H_2\over\sin\pi\alpha_2} &=&\int_0^{H_2^*} dw
\left ({H_2^* -w\over w}\right )^{\alpha_2}
\left ({L^* -w\over L^* + H_1^* -w}
\right )^{\alpha_1}  .
\label{H1}
\end{eqnarray}
Equations (\ref{L}--\ref{H2}) form a system of three equations
determinating the parameters of the transformation: $L^*$, $H_1^*$ and
$H_2^*$.  If the step heights are small compared to the distance
between them, we obtain the solution taking into account the main order
of the correction:
\begin{equation}
L^* = L - {H_2\over\pi} \ln {
e \chi_2 L\over H_2} + {H_1\over\pi}\left ( \ln { e \chi_1 L\over H_1}
+ {\pi\over\tan{\pi\alpha_1}} -{1\over\alpha_1}\right )\;,
\label{L*2}
\end{equation}
\begin{equation}
H_1^* =\left (1-{H_2\over\pi L}\right ){H_1\over\pi\alpha_1} \,,\;\;\;
H_2^* =\left (1+{H_1\over\pi L}\right ){H_2\over\pi\alpha_2} \,,
\label{H*2}
\end{equation}
where
\begin{equation}
\chi_{1,2} = \chi (\alpha_{1,2}), \;\;
\ln{\chi (\xi )} = \psi (2) -\psi (1-\xi ) +\ln{\pi\xi },
\label{chi}
\end{equation}
and $\psi (z)$ is the digamma function. Note, that the corrections are
nonanalytic for a small parameter $H_{1,2}/L\ll 1$. For the
rectilinear step profile $\chi (1/2) =2\pi$. In order to find the limit
$\alpha\rightarrow 0$ one has to put $H\rightarrow 0$, the
ratio $H/\alpha$ remaining as constant.

To see the kinetics of the steps, we determine the difference of the
solute concentration in the foot points of the steps $\Delta C = C(B_2)
- C(B_1)$.  Integrating (\ref{C}) for mapping (\ref{map},\ref{G2})
correct to the main order of $H/L$ leads to
\begin{eqnarray}
\Delta C &=& {J_1\over\pi D} \ln{\chi_1 L\over
H_1} - {J_2\over\pi D} \ln{\chi_2 L\over H_2}
\nonumber\\
&&-{J_1-J_2\over\pi^2 D L}\biggl( H_2\ln{\chi_2 L\over H_2} -
H_1\ln{\chi_1 L\over H_1} -{\pi
H_1\over\tan{\pi\alpha_1}}\biggr)
\nonumber\\
&&+{J_1 H_1\over\pi^2 D
L}\cdot {3\alpha_1 -1\over 2\alpha_1} +{J_2 H_2\over\pi^2 D L}\cdot
{3\alpha_2 -1\over 2\alpha_2}\;,
\label{DC2}
\end{eqnarray}
where $J_1$ and $J_2$ are full fluxes of the solute to the step risers
per unit length along the step. Note that $\Delta C$ is not a sum of
contributions going from two steps, and it contains cross-terms of
order of $H/L$ in the limit $H/L\rightarrow 0$. Two first terms of
(\ref{DC2}) present the diffusion field as a superposition of the field
going from two point sources. The rest terms are a correction
proportional to $H/L$, which takes into account the finite height of
the steps.

It is easy to see that the sign of the concentration difference depends
on the steepness of the steps. If the steps are identical, $H_1 =H_2
=H,\;\;\alpha_1 =\alpha_2 =\alpha$, and the fluxes are equal to each
other, $J_1 =J_2=J$, then (\ref{DC2}) takes the form
\begin{equation}
\Delta C ={J H\over\pi^2 D L}\cdot {3\alpha
-1\over\alpha}\;.
\label{=2} \end{equation}
For steep-sided steps, $\alpha > 1/3$, the concentration at the
foot of the leading step is higher and for gentle steps, $\alpha <
1/3$, it is quite the contrary. So for the steep-sided case, the
overtaking second step is approaching, and for the gentle steps,
the second is slow.

We have found the concentration difference in the limit of great
distances between steps. In the inverse limit, after merging, $L=0$,
there is a step with the summarizing height $2H$ and flux $2J$. The
$h$--dependence of the concentration difference between the points on
the riser of the step having joined: the point spacing from the foot
by the height $h$ and the foot has the form
\begin{equation}
\Delta C_0 =
{J\over D H}\left( h^* - {h\over\tan\pi\alpha}\right)\,,
\label{merge}
\end{equation}
where $h^*$ is defined by an equation
\begin{equation}
\pi\alpha h\csc\pi\alpha = 2 H B_{h^*/H^*}(1-\alpha ,1+\alpha ).
\label{eq}
\end{equation}
Here $B_z(a,b)$ is the incomplete beta function.  The expression
(\ref{merge}) shows that only for $\alpha\geq 1/2$ the foot
concentration is lesser than all over the riser. To determine the sign
of $\Delta C_0$ for identical steps one has to put $h=H$. There is some
$\alpha_0$ at which $\Delta C_0=0$, the difference being negative at
$0<\alpha<\alpha_0$, and being positive at $\alpha_0 <\alpha <1/2$.
The value of $\alpha_0$ is determined by the equation
\begin{equation}
{\pi\alpha_0\over 2\sin\pi\alpha_0}=
B_{\pi\alpha_0\over 2\tan\pi\alpha_0}
(1-\alpha_0 ,1+\alpha_0).
\label{a0}
\end{equation}
The numerical solution of (\ref{a0}) gives $\alpha_0\approx 0.37534$. It
means that the steps closing in can be merged at $\alpha >\alpha_0$.

\section{Diffusion nearby step train}
Let us pass over to consideration of the infinite step train shown in
fig.~2. It is formed by alternation of steps of height $H$,
terrace width $L$ and slope $A$ and of steps with parameters $h$,
$l$ and $\alpha$, respectively. The conformal mapping (\ref{map}) at
$\Gamma =\Gamma_\infty$,
\begin{equation}
\Gamma_\infty (w) = \prod_{k=-\infty}^\infty
{\left ({w-A_k^*\over w-B_k^*}\right )^{\alpha_k}}\,,
\label{gt}
\end{equation}
where
\begin{eqnarray}
&A_{2k}^* =L^* + P^* k\,,\; B_{2k}^* = L^* +h^*  + P^* k\,,\;
\alpha_{2k} =\alpha\,,
\nonumber\\
&A_{2k+1}^* =-H^* + P^* k\,,\;
B_{2k+1}^* = P^* k\,,\;  \alpha_{2k+1} =A,\nonumber\\
&P^*= \Lambda +\lambda,\; \Lambda = L^* +H^*,\; \lambda =l^* + h^*
\nonumber
\label{gt1}
\end{eqnarray}
transforms the area over the crystal into the upper half area of the
$W$-plain. Using the formula
\begin{equation}
\prod_{k=-\infty}^\infty
{a+ k\over b+ k}=
{\sin \pi a\over\sin \pi b}\,,
\label{S}
\end{equation}
we obtain
\begin{equation}
\Gamma (w) =\left [{\sin \kappa^* (w+H^*)
\over \sin \kappa^* w}\right ]^A
\left [{\sin \kappa^* (w-L^*)
\over \sin \kappa^* (w-L^* - h^*)}\right ]^{\alpha}
\,,
\label{Gt2}
\end{equation}
where $\kappa^* =\pi /P^*$. In analogy with the case of two steps to
dermine the unknown parameters $H^*$, $h^*$ and $L^*$, $l^*$ there
is a system of four equation. Two of these have the form
\begin{equation}
L = \int^{L^*}_0 dw
\left [{\sin \kappa^* (H^* +w)
\over \sin \kappa^* w}\right ]^A
\left [{\sin \kappa^* (L^* -w)
\over \sin \kappa^* (L^* + h^* -w)}\right ]^{\alpha}
\,,
\label{Lt}
\end{equation}
\begin{equation}
{H\over\sin \pi A} = \int^{H^*}_0 dw
\left [{\sin \kappa^* w
\over \sin \kappa^* (H^* -w)}\right ]^A
\left [{\sin \kappa^* (\Lambda -w)
\over \sin \kappa^* (\Lambda + h^* -w)}\right ]^{\alpha}
\,,
\label{Ht}
\end{equation}
and the second pair is obtained from (\ref{Lt},\ref{Ht}) by mutual
replacing of $H^*$, $L^*$, $A$ and $\Lambda$ by
$h^*$, $l^*$, $\alpha$ and $\lambda$.  At $H\ll L$, $h\ll l$ and
$|L-l|\ll L$, taking into account the first nonvanishing
correction, one obtains:
\begin{eqnarray}
L^* + l^* &=& L + l
+ {\pi (H - h) (L-l)\over L + l}\nonumber\\
&&- \left ({1\over\pi A} -{1\over\tan\pi A}\right ) H
- \left ({1\over\pi \alpha} -{1\over\tan\pi \alpha}\right ) h,\\
H^* &=& \left (1 - {\pi h (L-l)\over 2 (L+l)^2}\right )
{H\over \pi A}\;.
\label{lht}
\end{eqnarray}
Let $J$ É $j$ are the specific (per unite step length) solute fluxes to
risers of the odd and even steps respectively. The use of identity
(\ref{S}) gives an expression for the concentration difference at
points $B_2$ and $B_1$ in the form
\begin{eqnarray}
\Delta C &=& {J\sin\pi A\over\pi D H} \int_0^{H^*} dw
\left |\Gamma_\infty (w - H^* )\right |\;
\ln{\sin \kappa^* (l^* + w)\over\sin\kappa^* (H^* -w)}
\nonumber\\
&&-{j\sin\pi\alpha\over\pi D h} \int_0^{h^*} dw
\left |\Gamma_\infty (w + L^*)\right |\;
\ln{\sin\kappa^* (L^* + w)\over\sin\kappa^* (h^* -w)}.
\label{dct}
\end{eqnarray}
Integrating (\ref{dct}) with the accuracy mentioned above, we obtain
\begin{eqnarray}
\Delta C &=& {J\over\pi D} \ln{\chi_A (L+l)\over\pi H}
-{j\over\pi D} \ln{\chi_\alpha (L+l)\over\pi h}
\nonumber\\
&&+{J-j\over\pi D (L+l)}
\left ({H\over\tan\pi A} + {h\over\tan\pi\alpha}\right )
+{L-l\over 2 D (L+l)^2}
\nonumber\\
&&\times\left [
J\left ({1+3 A\over 2 A} H -{h\over\alpha}\right )
+j \left ({1+3 \alpha\over 2 \alpha} h -{H\over A}\right )
\right],
\label{dct1}
\end{eqnarray}
where $\chi_A =\chi (A)$, $\chi_\alpha =\chi (\alpha)$, and the
function $\chi (\xi)$ is defined by (\ref{chi}). Note that there are
cross-terms here the same as in the case of a pair of steps. They
occur owing to the influence of the geometry of all steps on the
diffusion field of a given step. Further the expression for the
concentration difference will be used for a linear stability analysis
of the step train.

\section{Stability of an equidistant train for growth from solution}
Let us consider small variations from a periodical sequence of
macrosteps. We shall assume that velocities of the steps are governed
by the values of concentration at the step feet at points $B_k$.
It can be realized for the layer growth when the overhanging peaks
don't arise, for example, owing to the Gibbs-Thomson effect. In this
case, the velocity of the whole macrostep is determined by the advance
velocity of the elementary layer in the ground floor. This velocity
depends on the solute concentration $C(B_k)$. The rest layers making up
the macrostep move over the lower one, their velocities being
controlled by this lower layer. Generally, the profile of the step
riser is determinated by the surface anisotropy and the concentration
field at the riser. It is supposed for simplicity that the riser
of step is rectilinear.

For an equidistant train of identical steps, the concentration at
points $B_k$ are equal and, therefore, the advance velocities of the
steps are equal too. When there are deviations from these position, the
concentration field and the velocities are varied. On the other hand,
the difference of the step velocities varies the step positions. The
presence of this back coupling can lead to an instability. The nearest
steps influence most strongly, so the perturbation doubling period of
the train has a maximum increment.  For the sequence of steps with
identical heights and slopes and at terraces $L$ and $l$, (\ref{dct1})
is reduced to
\begin{eqnarray}
\Delta C &=& {J-j\over\pi D} \left (\ln{\chi_A
(L+l)\over\pi H} + {2 H\cot\pi A\over L+l}\right )
\nonumber\\
&&+ {J +j\over D}\cdot {3 A -1\over 4 A}\cdot {H (L-l)\over (L+l)^2}.
\label{dct2}
\end{eqnarray}
The difference of the advance velocity of the steps with even and odd
numbers is given by
\begin{equation}
\Delta V = \beta\,\Delta C,
\label{kin}
\end{equation}
where $\beta$ is the kinetic coefficient of the step. On the other
hand, it can be expressed through a time derivative of the difference
$\Delta L =L-l$
\begin{equation}
\Delta V = {1\over 2}\,{d\Delta L\over dt}.
\label{dt}
\end{equation}
{}From (\ref{dct2}--\ref{dt}) taking into account $J-j= - \rho H\,\Delta
V /D$, where $\rho$ is the crystal density, we derive a differential
equation for $\Delta L$, which has the solution
\begin{equation}
\Delta L = \Delta L_0\exp(V t/S).
\label{exp}
\end{equation}
Here $\Delta L_0$ is the initial deviation of the period and $S$ is
characteristic path length of the train for doubling its period.
\begin{equation}
S = {4 A\over\pi (3 A -1)}
\left ({\pi D\over \beta\rho H}
+\ln{2\chi_A L\over\pi H} \right ){L^2\over H}\,.
\label{length}
\end{equation}
For positive values of $S$, the equidistant train is stable against
doubling period and values $S<0$ correspond to stable state. The first
term of (\ref{length}) is proportional to the ratio of the diffusion
rate $D/H$ to the kinetic rate $\beta\rho$. The second term gives an
increase of $S$, due to that the concentration near by the fast-speed
step is less.

The expression (\ref{length}) shows that the instability occurs only
for steps at the angle of inclination more than $\pi /3$.
The instability coefficient $L/S$ increases with an increase of the
step height. When the riser slope is controlled by the anisotropy of
the crystal face and it remains the same after merging, the development
of the morphological instability is going by means of the consecutive
merging of the steps in pairs. This way will occur until the step
separations are much less than the thickness of the diffusion
layer.

To see if this instability can be significant in realistic experimental
situations, we estimate the quatity of the instability length $S$ using
the values typical to crystal growth from solution. As an example, let
us consider the face $\{101\}$ of KDP and ADP crystals. Great
instability is observed at high normal rate of growth, when the mean
slope $p=H/L$ of the vicinal face with respect to the singular one is
significant. Let at $A=1/2$ the macrosteps consisting each of four
elementary layers, $H= 2\cdot 10^{-7}$ cm, have the distances between
them $10^{-5}$ cm. The kinetic coefficients of step for various advance
directions have values from $\beta\rho\sim 0.05$ cm/c to $\beta\rho\sim
0.3$ cm/c \cite{rash}.  Using $D =5\cdot 10^{-6}$ cm$^2$/c gives the
estimate that $S\sim (0.4\div 2)\cdot 10^5 L = (0.17\div 1)$ cm. This
quantity is quite observed under real growth conditions of KDP. The
estimation shows that when growing a large KDP crystal with the faces
of about dozens of centimetres at high growth rate \cite{Besp,P}, the
diffusion overlapping of the concentration field may lead to arising
large macrosteps. They can give rise to inclusion of the solution in
the crystal deteriorating optical quality of the crystal. For the face
$\{101\}$ of ADP crystal, where the kinetic coefficients of steps are
higher \cite{Ku}, the doubling length in various direction lies in the
interval $S\sim (0.05\div 0.25)$ cm.

The growth conditions considered above are close to the conditions
under which the morphological instability connected with a flow inside
the diffusion layer was observed \cite{Ch1,Ch2}, the wavelength was
being much more than the step separation. In the former case the
maximum instability occures for more large wavelength of the
perturbation than considered here.

It is possible that a simular diffusion interaction takes place when a
crystal like GaP is grown from a high-temperature solution \cite{Nish}.

\section{Growth from melt}
We now examine the heat transfer and train instability for various
relations between the thermal diffusion constants of the crystal $K_c$
and of the melt $K_m$.
\subsec{$K_c\ll K_m$}
If the heat conductivity of crystal is negligible compared to
the heat conductivity of melt, then the solution reduces to the
solution of concentration problem considered above. Thus, the train
stability is at $0<A<1/3$.

\subsec{$K_c = K_m\equiv K$}
For equal heat conductivities, for a moving pair of steps shown in
fig.~2, we have, accoding to (\ref{T}), an expression for the difference
of the undercooling $\Delta T =\left[ T_m -
T(B_2)\right] - \left[T_m - T(B_1)\right]$ in the form
\begin{eqnarray}
\Delta T &=&{q_1\over 2\pi K} \int_0^1
\hbox{d}t\;\ln\left| {L + t\;g_1 H_1
\over (t-1) g_1 H_1}\right|
\nonumber\\ &&+ {q_2\over 2\pi
K} \int_0^1 \hbox{d}t\;
\ln\left| {t\;g_2 H_2
\over L + g_1 H_1 + t\;g_2 H_2}\right|,
\label{T2}
\end{eqnarray}
where $g_{1,2}=\csc(\pi\alpha_{1,2})\,\exp(i\pi\alpha_{1,2})$, $q_1$
and $q_2$ are the heat flux from the steps per unit step length. At
$H_{1,2}\ll L$, we obtain

\begin{eqnarray}
\Delta T &=&
{q_1\over 2\pi K} \ln{e L\sin\pi\alpha_1\over h_1}
- {q_2\over 2\pi K} \ln{e L\sin\pi\alpha_2\over h_2}
\nonumber\\
&&  + q_1 {H_1\cot\pi\alpha_1\over 4\pi K L}
 - q_2 {H_2\cot\pi\alpha_2\over 4\pi K L}
 - q_2 {H_1\cot\pi\alpha_1\over 2\pi K L}.
\label{T22}
\end{eqnarray}
For identical steps and equal heat fluxes $q_1 =q_2 =q$, the expression
(\ref{T22}) reduces to
\begin{equation}
\Delta T =- {q\, H\cot\pi\alpha\over 2\pi K L}.
\label{T222}
\end{equation}
Thus, the undercooling, which determines the velocity of the second step
moving behind the leader is less than that near the first, $\Delta T <
0$, at $\alpha <1/2$. It means that the second step drops behind in
this case. Therefore an equidistant train of nonoverehanging steps is
stable against the step bunching. For overhanging steps, $\alpha >1/2$,
there is stable steady motion.

To find the quantity of the relaxation of perturbations we evaluate the
difference of the undercooling at points $B_2$ É $B_1$ in fig.~2
\begin{eqnarray}
\Delta T &=&{Q\sin\pi A\over 2\pi H K}\sum_{k=-\infty}^\infty
\int_{A_{2k}}^{B_{2k}}|\hbox{d}z|\;\ln\left|
{B_2 - z\over B_1 -z}\right|
\nonumber\\
&&
+{q\sin\pi\alpha\over 2\pi h K}\sum_{k=-\infty}^\infty
\int_{A_{2k+1}}^{B_{2k+1}}|\hbox{d}z|\;\ln\left|
{B_2 - z\over B_1 -z}\right|,
\label{=T}
\end{eqnarray}
where
\begin{eqnarray}
&A_{2k} =L + P k\,,\;\;
B_{2k} = L +h\hbox{e}^{-i\pi\alpha}
\csc\pi\alpha
 + P k\,,
\nonumber\\
&A_{2k+1} =-H\hbox{e}^{-i\pi A}\csc\pi A + P k\,,
B_{2k+1} = P k\,,
\nonumber\\
&P= L +l
+H\hbox{e}^{-i\pi A}\csc\pi A
+ h\hbox{e}^{-i\pi\alpha}\csc\pi\alpha.
\nonumber
\label{=Tn}
\end{eqnarray}
Here, $Q$ and $q$ are the specific (i.e. per unit step length) heat flux
from the odd and even step, respectively. Carring out the sum in
(\ref{=T}) gives
\begin{eqnarray}
\Delta T &=&{Q\over 2\pi K}
\int_0^1\hbox{d}t\;\ln\left|
{\sin\kappa\left[B_0 + t(B_1 - A_1)\right]
\over\sin\kappa t(B_1 - A_1)}\right|
\nonumber\\
&&
+{q\over 2\pi K}
\int_0^1\hbox{d}t\;\ln\left|
{\sin\kappa (1-t)(B_0 - A_0)\over
\sin\kappa\left[A_0 + t(B_0 - A_0)\right]}\right|,
\label{=T1}
\end{eqnarray}
where $\kappa =\pi/P$. At $H\ll L$, $h\ll l$ and $|L-l|\ll L$, taking
into account the first nonvanishing correction we obtain an expression
for the undercooling difference of the even and odd steps in the form
\begin{eqnarray}
\Delta T &=& {Q\over 2\pi K} \ln{2 e (L+l)\over\pi H\csc\pi A}
-{q\over 2\pi K} \ln{2 e (L+l)\over\pi h\csc\pi\alpha}
\nonumber\\
&&-{\pi (L-l)\over 8 K (L+l)^2}
\left({Q h\over\tan\pi\alpha} + {q H\over\tan\pi A}\right).
\label{=dT}
\end{eqnarray}
The difference of the step velocities is given by $\Delta V
=\beta\,\Delta T$, and the difference of the heat fluxes is given by
$Q-q = - H L_v\Delta V$, where $L_v$ is the latent heat per unit volume
of the crystal.  Supposing that the parameters of the even steps are
the same as those of the odd steps, we arrive at:
\begin{equation}
S = - {4\tan\pi A\over \pi^2}\left(
{2\pi K\over\beta L_v H} +
\ln{4 e L\sin\pi A\over\pi H}\right){L^2\over H}.
\label{ST}
\end{equation}
Expression (\ref{ST}) shows that the instability occurs only for
overhanging steps.
\subsec{$K_c\gg K_m$}
Solution for this case can be derived from the solution of the
problem of diffusion above the crystal by replacing
$\alpha\rightarrow -\alpha$, $D\rightarrow K$ É $J\rightarrow Q$.
One should also replace $\Delta C\rightarrow -\Delta T$, since
for negative values of the riser angle, the order of step movement is
inverse. Then the expression (\ref{=2}) for the undercooling difference
takes the form
\begin{equation}
\Delta T =- {Q H\over\pi^2 K L}\cdot {3\alpha+1\over\alpha}\;.
\label{=2T}
\end{equation}
Thus, if the heat transfers through the crystal, then the overlapping
thermal field always stabilizes the equidistant train against the
bunching. The characteristic relaxation length is given by
\begin{equation}
S_T=-S = {4 A\over\pi (3 A +1)}
\left ({\pi K\over \beta L_v H}
+\ln{2\chi_A L\over\pi H}
+{\pi\over\tan\pi A}-{1\over A}\right ){L^2\over H}\,.
\label{lengthT}
\end{equation}

\section{Summary and conclusion}
The stability areas of the riser slope are shown in fig.~3 for the cases
studied here. For the diffusion interaction and for heat transfer
through melt, the equidistant train is stable against bunching at
$0<A<1/3$ as it is shown in upper diagram. For the heat interaction, at
equal heat conductivities, the train of nonoverhanging steps is stable
at $0<A<1/2$ (middle diagram). Finally, there is the stability at the
heat transfer through crystal for all values of $A$, as it is shown in
the lower diagram.  Since the presence of diffusion interaction through
crystal stabilizes the equidistant train, one can expect stability
at all $K_m \leq K_c$ at least for nonoverhanging steps, $0<A<1/2$.

For the diffusion interaction at small step heights, according to
(\ref{length}) the doubling length is proportional to $p^2$. Thus,
each next merging occurs on the same path length, since the slope of
vicinal face is staing the same, when the steps are bunching.  When $H$
becomes of order of
\begin{equation}
\tilde H ={\pi D\over\beta\rho\ln\left[2\chi_A/(\pi p)\right]}\,,
\label{th}
\end{equation}
it will be $S\sim (L\ln1/p)/p$, i.e. the doubling length begins to
grow. As a result of doubling, the distance between steps can become
comparable with the thickness of the diffusion layer. In this case it is
needed to take into account the solution flow, that gives
stabilization of the train against the short wave fluctuations.

Here, we have addressed only the step kinetics of positions of the
steps. In general, besides this there are motions of the macrostep
form: both the profile and the shape of step line on the face.

Let us consider the stability of the train during dissolution and
melting.  Let for the case of growth, the difference of concentration
nearby two steps $\Delta C_g$ and the difference of the absolute
value of the advance velocity $\Delta V_g =\beta\Delta C_g$ have been
determined.  These results can be transformed to solve the dissolution
problem by changing the signs of the fluxes. So the concentration
difference during the dissolution $\Delta C_d =-\Delta C_g$. Since the
advance velocity during dissolution is more for lesser
concentration, we have $\Delta V_d =\Delta V_g$, i.e. the sign is the
same. During the dissolution the steps move in reverse direction, so if
they are moving away from each other during the growth, than they are
drawing together during the dissolution. It means that the stable train
for the case of growth is unstable for the case of dissolution. The
same is true for growth from melt.

In conclusion, we shall discuss the kinetics of elementary steps.
In this case instead of the diffusion equation one must use a
microscopic discription. For surface diffusion, it was shown
\cite{H} that the correction of the diffusion equation is
proportional to the ratio of relaxation times characterizing surface
diffusion and desorbtion. For the case of volume diffusion one can
expect that the effect considered here will take place as well. To
estimate the order of the absolute of $S$ it is possible to use the
expression derived here, but the sign determining the stability of
steady motion can be replaced.

\section*{Acknowledgements}
The author expresses his gratitude to Professors V.I.Bespalov and
A.A.Chernov for useful discussions.

\newpage
\thispagestyle{empty}
\section*{Figure captions}
Fig. 1. Transformation of the area in $Z$-plane occupied by solution
into upper half of the $W$-plane.\\

\noindent
Fig. 2. Conformal mapping for the infinite train of alternate steps.
The odd steps have the height $H$, the riser angle $\pi A$ and the
terrace with width $L$; the evens have respectevely: $h$, $\pi\alpha$
and $l$.\\

\noindent
Fig. 3. Diagrams of stability areas of equidistant train. The black
area in the upper diagram presents the riser angles of the stable train
during growth from solution and during growth from melt, the heat
transfering through the melt. The black area in the middle diagram
corresponds to stable motion at growth from melt, when the heat
conductivity of the crystal and the melt are equal, and in the lower
one -- when the latent heat transfer through the crystal only. For
dissolution and melting the stable states correspond to light areas. At
the foot of the figure, the profiles of step at $A=1/4;\;1/2$ and 3/4
are shown.
\end{document}